\documentclass[12pt]{article}
\usepackage{setspace}
\usepackage{color}
\usepackage{graphicx}
\usepackage{cite}
\usepackage{amsmath}
\usepackage{xr}
\usepackage{floatpag}
\usepackage{times}
\usepackage{float}
\usepackage{fancyhdr}

\fancypagestyle{plain}{
  \fancyhf{}
  \fancyfoot{}
  
  \fancyhead[R]{\thepage}
 }

\fancypagestyle{title}{%
  \fancyhf{}
  \fancyfoot{}
  
  \fancyhead[L]{\framebox{\emph{Journal of the Royal Society Interface} {\bf 12}, 20150235 (2015) $|$ doi:10.1098/rsif.2015.0235}}
}

\newenvironment{rsiabstract}{%
\begin{quote}\bf}{
\end{quote}
}

\renewcommand\figurename{{\bf Fig.}}

\title{Regularity Underlies Erratic Population Abundances in Marine Ecosystems}

\author{
Jie Sun$^{1,4}$, Sean P. Cornelius$^{2,3,4}$, John Janssen$^{5}$,\\
Kimberly A. Gray$^{6}$, Adilson E. Motter$^{4,7,*}$\\
\\
\footnotesize{$^{1}$Department of Mathematics, Clarkson University, Potsdam, NY 13699, USA}\\
\footnotesize{$^{2}$Center for Complex Network Research and Departments of Physics, Northeastern University} \\
\footnotesize{Boston, MA 02115, USA}\\
\footnotesize{$^{3}$Channing Division of Network Medicine, Brigham and Women’s Hospital, Harvard Medical School,}\\
\footnotesize{Boston, MA 02115, USA}\\
\footnotesize{$^{4}$Department of Physics and Astronomy, Northwestern University, Evanston, IL 60208, USA}\\
\footnotesize{$^{5}$School of Freshwater Sciences, University of Wisconsin, Milwaukee, WI 53204, USA}\\
\footnotesize{$^{6}$Department of Civil and Environmental Engineering, Northwestern University, Evanston, IL  60208, USA}\\
\footnotesize{$^{7}$Northwestern Institute on Complex Systems, Northwestern University, Evanston, IL 60208, USA}\\
\footnotesize{$^\ast$To whom correspondence should be addressed; E-mail:  motter@northwestern.edu}
}

\date{}

\begin{document} 
\maketitle 
\thispagestyle{title}

\pagestyle{plain}
\vspace{-1cm}
\begin{rsiabstract}
The abundance of a species' population in an ecosystem is rarely stationary, often exhibiting large fluctuations over time. 
Using historical data on marine species, we show that the year-to-year fluctuations of population growth rate obey a well-defined double-exponential (Laplace) distribution. This striking regularity
allows us to devise a stochastic model despite seemingly irregular variations in population abundances. The model identifies the effect of reduced growth at low population density as a key factor missed in current approaches of population variability analysis and without which extinction risks are severely underestimated. The model also allows us to separate the effect of demographic stochasticity and show that single-species growth rates are dominantly determined by 
stochasticity common to all species.  This dominance---and the implications it has for interspecies correlations, including co-extinctions \newline ---emphasizes the need of ecosystem-level management approaches to reduce the extinction risk of the individual species themselves. 
\end{rsiabstract}


\baselineskip24pt

\clearpage
\section{Introduction}
Assessment of extinction risk and biodiversity loss is a central problem in ecology, which has direct implications for ecosystem management practices \cite{Lessard2005} and policies for the exploitation of natural resources \cite{Price2007}. It is estimated that currently up to 0.1\% of all known species go extinct every year, which is over one thousand times above the background extinction rate observed in fossil records \cite{Lawton1995}. Whether caused by habitat degradation, interspecies competition,  climate change, overexploitation, or the introduction of exotic species, the majority of such extinction events have not been anticipated. Existing inventories of the global conservation status, such as the IUCN Red List \cite{Mace2008}, are believed to include only a fraction of all endangered species and, even for those, significant uncertainty remains on their actual extinction risk. The central difficulty is that wild populations generally do not exist in a steady state from which minute deviations or trends could be detected. Instead, they tend to exhibit fluctuations over time \cite{McArdle1990, Gaston1994, MacArthur1955, Pimm1988, Pauly1998}.

Temporal variations in population abundance may be caused by species interactions, environmental changes, migration patterns, intrinsic nonlinearities, and/or human exploitation \cite{Anderson2008,Post2002,Beninca2008}, and are sometimes sufficiently irregular to be regarded as stochastic. Such irregular time dependence, when combined with unavoidably imperfect sampling of the population, poses considerable challenges for population forecast. Significant previous research has focused on determining the frequency composition (so called ``noise colour'') of such fluctuations and their correlations with the environment \cite{Halley1999, Vasseur2004}. Despite their immediate implications for sustainable ecosystem exploitation and management, much less understanding has been generated about the factors that influence the growth rate of individual species.  

Perhaps not surprisingly, historically there has been a large number of both false positives and false negatives in the assignment of extinction status and extinction risk.  For instance, the New Zealand's bird {\it takah\={e}}, which was considered extinct by the end of the 19th century, was rediscovered in the wild 50 years later \cite{DelHoy1996}. This is one of now many known examples of a Lazarus taxon \cite{Shuker2002}, in which a sparse population passed undetected for an extended period of time. The \textit{passenger pigeon} in North America, on the other hand, went from an abundant population to functional extinction in less than 20 years---which then  led to actual extinction in the early 20th century \cite{Schorger2004}. Taken together, the picture that emerges is one in which the survival of a species appears to depend very subtly on the species' own abundance \cite{Inchausti2003}. This picture is further complicated by the possibility of co-extinctions \cite{Srinivasan2007, Allesina2009}, whose actual role beyond directly dependent species (such as predator-prey and parasite-host) remains elusive \cite{Koh2004, Saavedra2011, Sahasrabudhe2011}. Compared to the terrestrial case, extinctions of marine species caused by habitat loss or human exploitation have been rare, in part due to stabilizing effects such as changes in fish catchability at low population densities. Yet, the pace of marine defaunation is likely to accelerate dramatically as the strength and scope of human impact on marine ecosystems grow \cite{McCauley2015}. This underscores the need for a quantitative understanding of the growth dynamics of marine populations, including extinction risks.

\section{Results}
Here, we examine the dependence of growth rate on population abundance and stochastic factors, both when only the temporal variability of individual species is observed and when the dynamics of the entire ecosystem is taken into consideration.  We base our analysis on data of the Large Marine Ecosystems (LME) portion of the Sea Around Us Project (http://www.seaaroundus.org) \cite{Pauly2007}, which is a global-scale database on species abundance in $66$ marine ecosystems. For each ecosystem, the data consist of the annual quantities (in tonnes) of its 12 most abundant species caught by fisheries over the period $1950$ to $2006$. Such landing data is a widely used tool for making inferences about marine populations\cite{Jackson2001,Lajus2005,Worm2009,Costello2012}, even if such use is occasionally controversial \cite{Pauly2013}, as factors other than population abundance can influence commercial catches. We will show, however, that our results hold true for marine stock assessments, which integrate research surveys and catch data with independent information (such as mortality rates and population size/age structure) to obtain a more accurate estimate of population abundance (supplementary material, \textit{Analysis of Stock Assessments}). Therefore, we present our analysis based on the more widely-available landing data \cite{Pauly2007}, followed by supplementary validation using stock assessment data.

For each species $i$ in a given ecosystem, we assume that the annual abundance in year $t$ can be approximated---up to a scaling factor---by the reported landing, denoted by $x^{(i)}_t$, although we will show that our results do not depend critically on this assumption (supplementary material, section 1, figures S1-S3). Our object of study is the year-to-year growth rate,  $r^{(i)}_t = \ln \big ( x^{(i)}_{t+1} / x^{(i)}_t \big )$. To avoid ill-defined log functions associated with zero landings, we add $1$ to each data point, so that the minimum value of $x^{(i)}_t$ is  $1$. Without any further assumptions, the year-to-year change in the species' population abundance can always be written as
\begin{equation}\label{eq:1}
	x^{(i)}_{t+1} =  e^{\bar{r}^{(i)} + \sigma^{(i)} \xi^{(i)}_t}x^{(i)}_t,
\end{equation}
where $r^{(i)}_t= \bar{r}^{(i)} + \sigma^{(i)} \xi^{(i)}_t$ represents the growth rate in year $t$ decomposed into the average $\bar{r}^{(i)}$ and standard deviation $\sigma^{(i)}$ of the growth rate (calculated over the entire time series) and a time-dependent factor $\xi^{(i)}_t$ at time $t$. The term  $\xi^{(i)}_t$ is thus the normalised growth rate fluctuation. For example, Eq.~\ref{eq:1} reduces to the classical Ricker model \cite{Ricker1954} if $r^{(i)}_t$ is a deterministic decreasing linear function of the  abundance $x^{(i)}_t$ which, as shown below, does not hold true for the marine ecosystems we consider. In contrast with previous studies, here we will make no a priori assumptions on the growth rate, instead deriving its properties directly from the data. 

Figure \ref{stats_fig} presents empirical properties of $x^{(i)}_t$ for all $66$ ecosystems we consider. We first note that while the autocorrelation of the (log) population abundances  $\ln x^{(i)}_t$ is significant and close to one for successive years, as expected (figure~\ref{stats_fig}A), the autocorrelation of the corresponding growth rates is comparatively low, with a majority of species having autocorrelation less than 0.2 in magnitude. This surprising observation indicates that the time-dependent component of the growth rate, $\xi^{(i)}_t$, can be regarded as a random variable drawn from an appropriate distribution that is nearly stationary. The data show that, to an excellent approximation, this distribution is given by a double-exponential function
\begin{equation}\label{eq:2}
	F_1(\xi)=\frac{1}{\sqrt{2}}\exp(-\sqrt{2}|\xi|), 
\end{equation}
also known as the Laplace distribution (figure~\ref{stats_fig}B-C). This is itself an important, novel finding, which further allows us to devise a statistical model, as discussed below. We have verified that our addition of 1's to the zero-landing years does not affect this distribution, which incidentally also governs the fluctuations of growth rate for the population abundance of each ecosystem as a whole (figure~\ref{stats_fig}D). In addition, we have rigorously confirmed the fit of the individual species' growth fluctuations to the Laplace distribution versus the null hypothesis of a normal distribution, according to standard goodness-of-fit tests such as the Kolmogorov-Smirnov test and Akaike information criterion. These tests show that the former distribution is a significantly more plausible explanation than the latter distribution for the growth rates of a large majority of species (supplementary material, figure~S4).

Having established that the normalised growth rate fluctuations obey a stationary stochastic process described by a parameter-free Laplace distribution,  we can propose Eqs.~\ref{eq:1}-\ref{eq:2} themselves as a stochastic model of population abundance.  This model is expected to be appropriate away from the {\it floor} abundance $x^{(i)}_t=1$. However, the growth rate at the floor is much smaller than at any other abundance, with probability  $89\%$ of being zero, as shown in the inset of figure~\ref{stats_fig}C. Naturally, because abundance is measured in integer units of landing tonnes, $x^{(i)}_t=1$ only indicates that the population is very low but not necessarily that it is zero. For this reason (and possibly due to migration and sample biases), the apparent ``extinctions'' considered here are local in space and in time. Indeed, the populations generally recover to detectable abundances in the systems under consideration. However, they do so more slowly than predicted by the overall growth rates. This reduced growth rate at low population density is consistent with the Allee effect (or depensatory population dynamics), which is a scenario previously observed for a number of species in diverse ecosystems \cite{Kramer2009}. Further analysis would be needed to conclusively address that particular issue in this case, which falls outside the scope of this paper. It should be noted, nevertheless, that the apparent reduced growth rate at low population abundances is counter to most existing models, including the classical Ricker model \cite{Chen2002}. A recent study of exploited marine species stocks included in the  RAM Legacy Stock Assessment Database \cite{Ricard2012} indicates that this missing element is in fact the probable explanation for the slow recovery of depleted stocks when compared with predictions from models commonly used in fisheries management \cite{Keith2012}.

We incorporate the floor effect into our model by taking the probability distribution of $\xi^{(i)}_t$ to be
\begin{equation}\label{eq:3}
F_2(\xi)=
\begin{cases}
F_1(\xi), &\mbox{ for } x^{(i)}_t>1,\\[8pt]
\begin{aligned}
(1-p^{(i)}) \times \delta(\xi-\xi^{(i)}_c)
+\; p^{(i)} \times 2F_1(\xi),
\end{aligned} &\mbox{ for }  x^{(i)}_t=1 \mbox{ and }  \xi\ge 0,
\end{cases}
\end{equation}
and $F_2(\xi)=0$ otherwise. The parameter $p^{(i)}$ is  a measure of the recovery probability once the species reaches the floor abundance, $\delta$ is the Dirac delta function, and $\xi^{(i)}_c=-\bar{r}^{(i)}/\sigma^{(i)}$ is defined so that $r_t^{(i)} = 0$ when $\xi_t^{(i)} = \xi^{(i)}_c$ (for simplicity in equation (3), we  used the approximation $\xi^{(i)} \ge 0$ instead of the exact condition  $\xi^{(i)}\ge \xi^{(i)}_c$, since $\bar{r}^{(i)}$---and hence $\xi^{(i)}_c$---is typically close to zero). Here, the average growth rate $\bar{r}^{(i)}$ and standard deviation $\sigma^{(i)}$ are estimated for $x^{(i)}_t>1$ and the recovery probability  $p^{(i)}$ is estimated for $x^{(i)}_t=1$ ({\it Methods}, Parameter estimation). The model  defined by Eqs.~\ref{eq:1}-\ref{eq:3} offers projections on future population abundance based on past abundance, which in turn can be used to assess risk of {\it pseudo-extinctions}, defined as crossings below a given fraction $\theta$ of the historical maximum population.

Figure \ref{model_validation_fig} validates our model against empirical data ({\it Methods}, Model predictions and validation). It shows, in particular, that the floor effect is crucial for the excellent agreement found between the pseudo-extinction predicted and the ones actually observed over the same period (figure~\ref{model_validation_fig}A). We have tested that the heightened agreement with the empirical data holds regardless of the exact value used for the floor abundance (which also represents the limit of resolution in the data), insofar as it is not too large (figure~\ref{alt_floor_fig}). This is significant because current approaches for population variability analysis, including specialised commercial software, generally do not account for the exceptional case of growth at very low population densities (even though minimal viable population sizes are often assumed \cite{Shaffer1981}). Neglecting the floor effect not only mis-predicts pseudo-extinctions observed in the past, but also tends to severely underestimate the risks of future pseudo-extinctions (figure~\ref{model_validation_fig}B). For example, the popular deepwater redfish ({\it Sebastes mentella}) in Baffin Bay appears to have recovered from very low population abundances, but historical data indicate that when the abundance of this species approaches zero it has a high probability of remaining extremely low for extended periods (figure~\ref{model_validation_fig}C). Neglecting this reduced growth leads to a significant underestimation of the pseudo-extinction risk. This underestimation is even more pronounced for a number of other species, such as the flatfishes  (Pleuronectiformes spp.) in the Antarctic (figure~\ref{model_validation_fig}D). Interestingly, for some species neglecting the floor effect may actually lead to an overestimation of the pseudo-extinction risk. For example, the blue mussel ({\it Mytilus galloprovincialis}) in the Iberian Coast reached the floor abundance in the year 2004,  but started recovering immediately (figure~\ref{model_validation_fig}E). Because the species exhibited positive growth the only time it reached the floor,  predictions that take this into account naturally lead to an estimate of pseudo-extinction risk that is smaller and more reliable than predictions that do not.

Central to our analysis is the fluctuation of the growth rate, modeled as a stochastic term $\xi^{(i)}_t$. But what is the relation between the stochasticity of different species $i$ in the same ecosystem? This question can be addressed by making the ansatz that the $\xi^{(i)}_t$ are not independent but are instead drawn from an appropriate \textit{joint} distribution in which the (marginal) distributions for the individual species follow Eq.~\ref{eq:3} while having correlation $\alpha^2$ with those of the other species ({\it Methods}, Estimation of common stochasticity). We then consider a range of values of $\alpha$, which represents the portion of stochasticity common to all species, and analyse the integrated impact on the population abundance of each ecosystem as a whole. To give all the species comparable weight, we focus on the average over the individual species' population weighted by the inverse of their average abundance (as in figure~\ref{stats_fig}D). The occurrence of pseudo-extinctions is systematically underestimated when stochasticity is dominantly species-specific, as illustrated in figure~\ref{common_stoch_fig}A for $\alpha=0$. In fact, the agreement of the model with empirical data is significantly better when stochasticity is taken to be dominantly common to all species (figure~\ref{common_stoch_fig}B), and the agreement becomes excellent for $\alpha=0.8$ (figure~\ref{common_stoch_fig}A). 

This corroborates the conclusion that variations in growth rate are largely synchronised within each ecosystem. We speculate that this synchronisation is partially
rooted in external, environmental fluctuations, which are known to play a crucial role in the dynamics of terrestrial populations \cite{Post2002,Blasius1999}. In marine systems, likely environmental fluctuations include the El Ni{\~n}o Southern Oscillation, the North Atlantic Oscillation, and riverine flood pulses. For fishes, the impacts of these fluctuations can be both direct and indirect---such as via externally-driven changes in the lower food web or via ecosystem regime shifts \cite{Scheffer2012}. Other potential sources of synchronized growth fluctuations include interspecies interactions as well as correlations with human activity---for example shifts in overall fishing effort due to weather or changes in regulations. But regardless of the source of the common stochasticity observed here, the direct implication is an increased risk of the otherwise unlikely concurrent collapse of multiple species.  \\

\section{Discussion} 

Our findings should be compared with the case studies of the American breeding bird populations \cite{Keitt1998} and Hinkley Point's fish community \cite{Storch2007}, for which growth rates have been analyzed. For both systems the aggregated non-normalised growth rates were found to follow power-law distributions, but independent analysis of some of these data has shown that after rescaling (by the species standard deviation) to a variable equivalent to the normalised growth rate fluctuation $\xi^{(i)}_t$, the distribution becomes normal \cite{Allen2001}, which corroborates the conclusion that the growth rate distributions of individual species are short-tailed. This is consistent with the previous analysis of 544 long-term time-series from the global population dynamics database \cite{Halley2002},  including both aquatic and terrestrial populations, which demonstrated that the abundances of most species are either lognormally distributed or shorter-tailed than lognormally.  
If one neglects the (significant) year-to-year correlations in abundance, lognormal distributions for individual population abundances imply normal distributions for individual species' growth rates. 

The results presented here, on the other hand, show that normalised growth rate fluctuations follow Laplace distributions and this is confirmed to remain true for individual species in the ecosystems we consider (supplementary material, figure~S4). Had we not normalised the growth rates to eliminate heterogeneity across species, the resulting aggregated non-normalised growth rate fluctuations would be more fat-tailed than what we observe, without necessarily revealing a simple scaling behaviour (supplementary material, figure~S5). Importantly, we have verified that the observed Laplace distributions for the normalised growth rates are not artefacts of the landing data used here as a proxy for population abundance,  remaining valid for stock assessment data. This is demonstrated in figure~S6 (supplementary material) for the RAM Legacy Stock Assessment Database \cite{Ricard2012}, which is the most complete marine stock assessment catalogue available. The estimates of population abundance therein were acquired under controlled settings, using a variety of methodologies designed to avoid systematic biases, and in a variety of ecosystems, which substantiates the conclusion that Laplace statistics underlie the growth of marine populations in general. Note that different types of growth rate distributions are consistent with the lognormally-distributed abundances observed in previous studies; by Eq.~\ref{eq:1}, the (log) population abundance is (up to a constant) the sum of the growth rates in all previous years. Thus, by the Central Limit Theorem, \emph{any} growth rate distribution (including Laplace) will eventually lead to normal distributions for the log abundances, provided the growth rates are independent and identically distributed with finite variance. However, direct analysis of the marine population abundances in this study reveals that their distributions are no closer to lognormal than to power laws (supplementary material, figure~S7). Altogether, our results are significantly different from those suggested by previous studies, and they do not follow from existing ecological models. 

Our demonstration that the growth rates of marine species are governed by the Laplace distribution, which has a heavier tail than a normal distribution with the same standard deviation, has important implications for the analysis of empirical data.  Because the likelihood of pronounced fluctuations is larger than expected from normal distributions, large short-term fluctuations (over the period of few years) are not necessarily a sign of abnormality as they may well be a natural property of the system. This, combined with the observed stickiness to the floor abundance, poses additional challenges to the identification of abnormal population dynamics. In particular, we have shown that neglecting the floor effect alone already leads to substantial underestimation of pseudo-extinction risks. Finally, since Laplace distributions have been previously identified in the growth of companies \cite{Stanley1996}, our study establishes a new parallel between ecological and socio-economical networks, both of which are characterized by growth and competition in the presence of limited resources. As such, our results may also provide new insights into common stability mechanisms that govern otherwise disparate systems---as recently proposed in ref.~\citen{Haldane2011}. 

\section{Methods}
\noindent \textbf{Parameter estimation.\ } To estimate the model parameters from a given time series of population abundance, we disregard the initial consecutive $1$'s, if any, as they often correspond to years for which no data are available. 
Using the resulting time series $\{x_t\}$, we compute the associated growth rate time series $\{r_t\}$ from the definition $r_t = \ln x_{t+1} - \ln {x_t}$, where we now omit the superscript $(i)$ for simplicity (a convention also adopted in the figures). The parameter estimation depends on whether or not we consider the floor effect in the model. For the model without floor effect (Eqs.~\ref{eq:1}-\ref{eq:2}), $\bar{r}$ and $\sigma$ are simply the sample mean and standard deviation of $\{r_t\}$.
For the model with floor effect (Eqs.~\ref{eq:1}-\ref{eq:3}), we have
$	\bar{r} = \frac{1}{|I_1|}\sum_{t\in I_1}r_t$,
$	\sigma = \sqrt{\frac{1}{|I_1|-1}\sum_{t\in I_1}(r_t-\bar{r})^2}$,
and  $p = 1 - \frac{|I_{00}|}{|I_0|}$,
where 
$I_1=\{t|x_t>1\}$, $I_0=\{t|x_t=1\}$, $I_{00}=\{t|x_t=1 \mbox{~and~} x_{t+1}=1\}$, and $|\cdot|$ denotes the number of elements in the~set. \\

\noindent \textbf{Model predictions and validation.\ } For each time series $\{x_t\}$ of length $T$, we use its first $T_0$ data points as the training set for parameter estimation and the remaining $T-T_0$ data points for validation. We calibrate our model both without and with floor effect, and simulate both variants for $1,000$ independent runs. Each run starts with the same initial condition $x_{T_0}$ and is computed for a total of $T-T_0$ time steps, with $x_t$ reset to $1$ whenever its estimated value is below $1$. For both the emprical and simulated data, we calculate the pseudo-extinction risk as the probability that $x_{t}\leq\theta{\cal X}$, where $\theta\in \left[\frac{1}{\cal X},1\right)$ is the pseudo-extinction threshold and ${\cal X}=\max_{1\leq t\leq T_0}x_t$. The special case of pseudo-extinction risk at the floor level is defined as the fraction of years for which the population abundance is at the floor (i.e., $x_{t}=1$). In this study, $T=57$ years and we choose $T_0=45$ years. \\

\noindent \textbf{Estimation of common stochasticity.\ } We investigate the existence of common stochasticity among all species considered in each ecosystem by extending our model as follows. In a given year $t$, for those species that do not remain at the floor abundance (according to Eq.~\ref{eq:3}), we draw a vector $\vec{\xi_t} = \left (\xi^{(i)}_t \right )$ of growth fluctuations from a multivariate Laplace distribution \cite{Kotz2001}
characterised by (vector) mean 0, (vector) variance 1, and normalised covariance matrix $\Gamma$. We set the diagonal elements of $\Gamma$ identically to $1$ and the off-diagonal elements identically to $\alpha^2$, where $\alpha$ is a parameter ranging from $0$ to $1$. The resulting growth fluctuations for the individual species are distributed according to Eq.~\ref{eq:2} but the correlation coefficient is $\alpha^2$ between $\xi^{(i)}_t$ and $\xi^{(j)}_t$ for any pair $i \neq j$. As such, the parameter $\alpha$ can be interpreted as  the proportion of common stochasticity. The value of $\alpha$ that best represents the empirical data is determined in figure~\ref{common_stoch_fig}. \\

{\noindent \bf Competing interests.\ } The authors declare that they have no competing interests. \\

{\noindent \bf Funding.\ } This study was supported by the U.S.\ National Oceanic and Atmospheric Administration under Grant No.\ NA09NMF4630406. \\

{\noindent \bf Authors' contributions.\ } 
All authors participated in the design of the research; 
JS and SPC performed data analysis and simulations; 
JS, SPC, and AEM wrote the manuscript; 
all authors commented on the manuscript and gave final approval for publication. 
JS and SPC contributed equally to this work.

\newpage
\clearpage
\pagestyle{plain}
{
\renewcommand{\markboth}[2]{}

}

\newpage
\clearpage
\begin{figure*}[h!]
\begin{center}
\includegraphics*[width=0.95\textwidth]{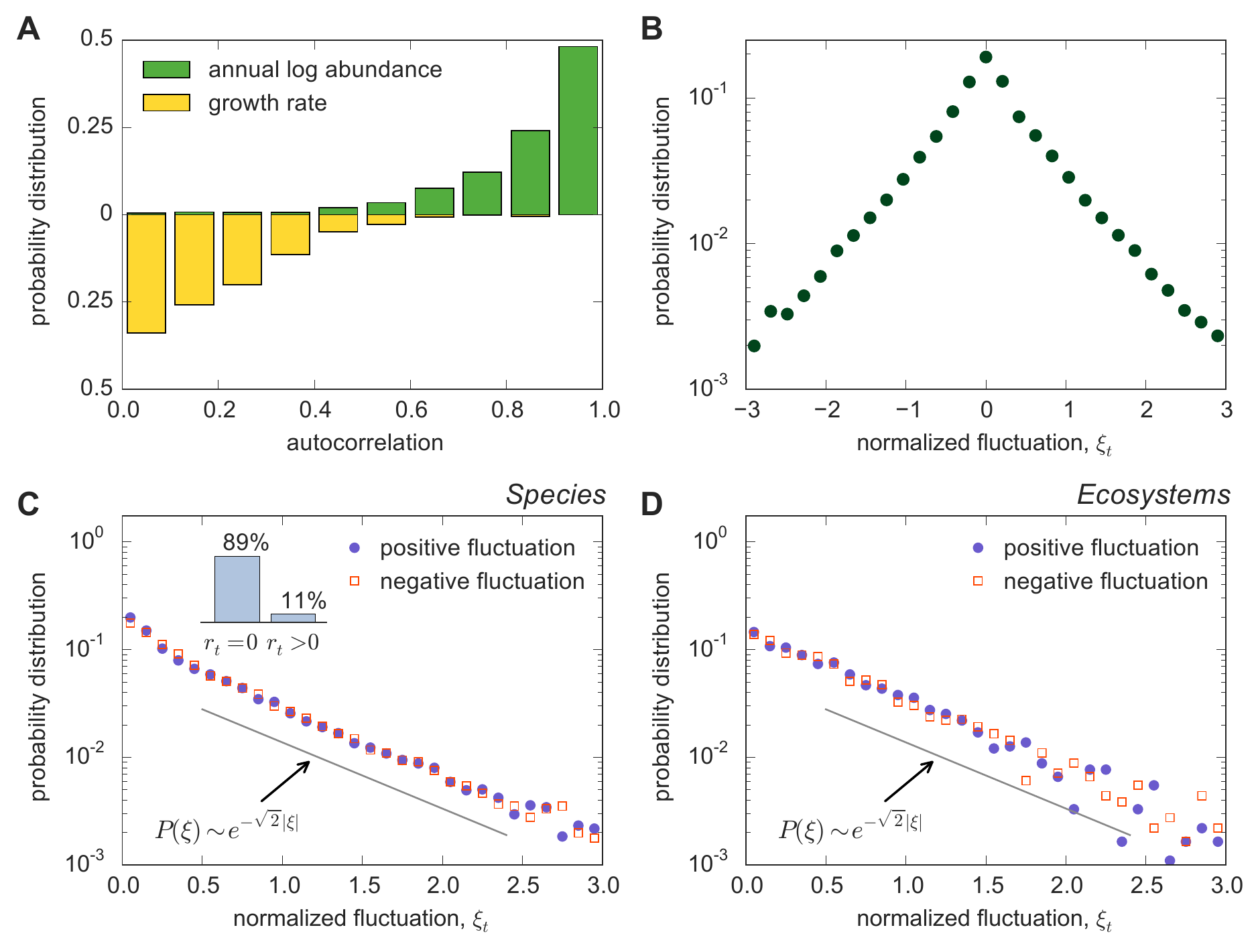} 
\end{center}
\caption{\label{stats_fig}
{\bf Statistical properties of population growth rate 
derived from the 
Large Marine Ecosystems dataset.}
\,
(\textbf{A}) Distribution of the absolute values of the lag-1 autocorrelation of the log-transformed annual population abundance $\{\ln {x_t}\}$ (upper part) and of the growth rate $\{r_t\}$ (lower part). For most species, $\{\ln {x_t}\}$ exhibits significant autocorrelation whereas $\{r_t\}$ does not (\textbf{B}) Normalised growth rate fluctuation $\xi_t$ for individual species, showing the double-exponential ``tent'' shape of the distribution.(\textbf{C}) Same as in panel~B, where the distribution of positive fluctuations $P(\xi_t|\xi_t\geq0)$ (solid symbols) and the distribution of the magnitudes of negative fluctuations $P(-\xi_t|\xi_t\leq0)$ (open symbols) are displayed separately to show the symmetry. Both distributions follow an exponential function $P(\xi)\sim\exp(-\lambda|\xi|)$, with $\lambda\approx\sqrt{2}$. The inset shows the probabilities $P(r_t=0|\ln {x_{t}}=0)$ and $P(r_t>0|\ln {x_{t}}=0)$. When the abundance of a species reaches the floor level  $\ln {x_{t}}=0$, it has a high probability (89\%) of staying at that level the following year, which reveals a strong relation between low population and nearly zero growth rate. We note that, while our addition of 1's to the reported populations will tend to deflate the apparent magnitude of the growth rate at small population abundances when calculated using Eq.~\ref{eq:1}, the stickiness to the floor reported here is a property of the raw data and is not influenced by this transformation. (\textbf{D}) Same as in panel~C at the ecosystem level,  where the `population abundance' of an ecosystem is taken to be the weighted average over all of its 12 recorded species, with the weights defined by the inverse of the average abundance of the species. The exponent $\sqrt{2}$ is within the $95\%$ confidence interval of the linear regressions.
} 
\end{figure*}

\begin{figure*}[h!]
\begin{center}
\includegraphics*[width=0.95\textwidth]{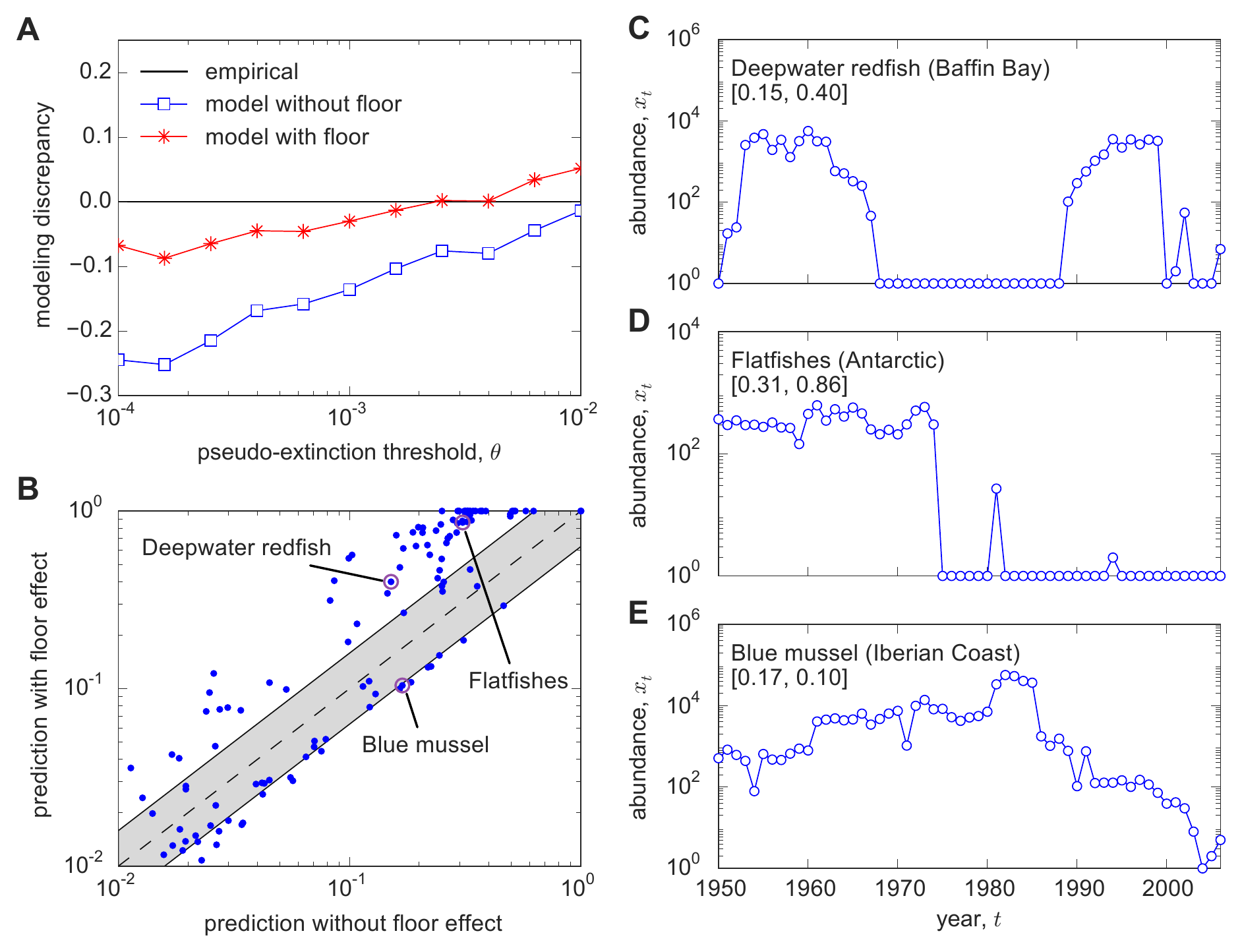} 
\end{center}
\caption{\label{model_validation_fig}
{\bf Model validation and predictions for individual-species pseudo-extinction risk.}
\,
(\textbf{A}) Model discrepancy as a function of the pseudo-extinction threshold, $\theta$, for a threshold range below $1\%$ of the maximum population. The curves show the average log-distance between the pseudo-extinction risk in the empirical data and that predicted by the model without floor (squares) and the model with floor (stars). The model is parameterized over the period 1950-1994, and the predictions are shown for the period 1995-2006. The curves correspond to averages over all recorded species of the ecosystems under consideration. (\textbf{B}) The risk of  pseudo-extinction at the floor level, projected for the period 2007-2021 for the model without and with floor parameterized over the full period 1950-2006 of available data. The higher incidence of points in the upper right corner above the shaded reason along the diagonal indicates that  the pseudo-extinction risk is grossly underestimated if the floor is not explicitly accounted for (note the logarithmic scale).  (\textbf{C} to \textbf{E}) Examples of empirical time series for individual species (marked dots in panel~B) with zero-landing years leading to large systematic discrepancies between the pseudo-extinction risk predicted by the model without  floor (first number) and with floor (second number). Pseudo-extinctions are defined as crossings of population abundance below the given threshold $\theta$ (measured relative to the historical maximum), while pseudo-extinctions at the floor level are operationally defined as the extreme case in which populations reach the floor abundance $x^{(i)}_t=1$. The associated risk is quantified as the fraction of years below threshold or at the floor, respectively. See \textit{Methods} for details on model validation and predictions.
}
\end{figure*}

\begin{figure*}[h!]
\begin{center}
\includegraphics*{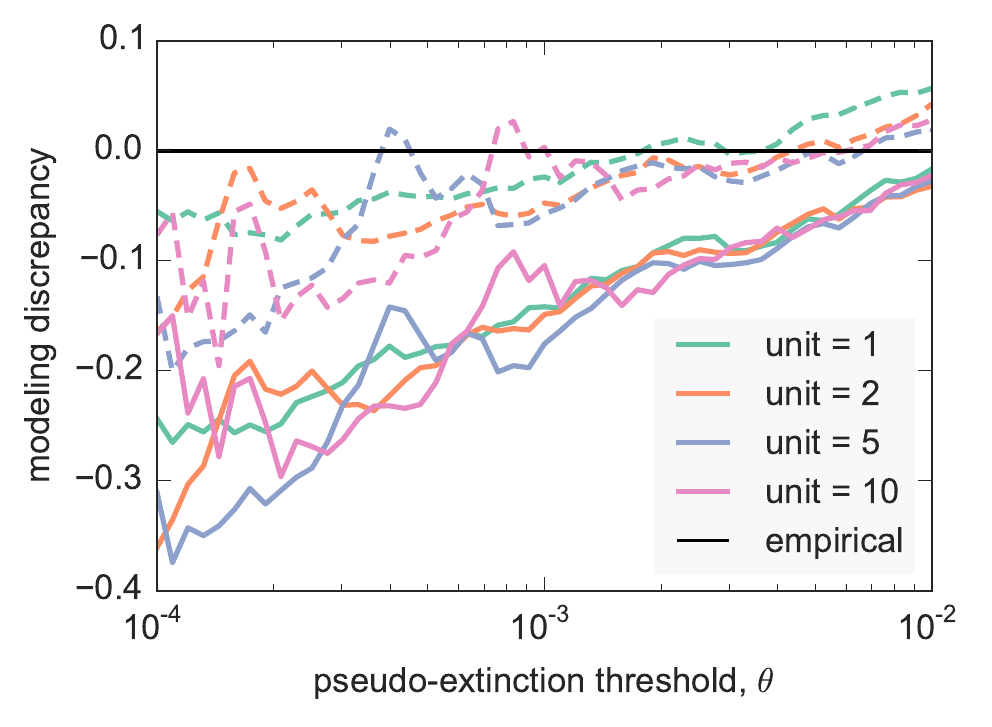} 
\end{center}
\caption{
{\bf Model validation for alternate floor values and levels of discretization.} Counterpart to figure~\ref{model_validation_fig}A for different values of the ``floor'' abundance (measured in tonnes), where dashed (continuous) lines indicate the model discrepancy with (without) floor. Incorporating the observed tendency for already low population abundances to remain low significantly increases the agreement with the empirically-observed pseudo-extinction risk for low threshold $\theta$, regardless of the precise value used to implement the floor effect. For each simulated value of the floor, it is assumed that the population abundances are not resolved below the floor level and that populations are only resolved to the nearest multiple of this level.
}\label{alt_floor_fig}
\end{figure*}

\begin{figure*}[h!]
\begin{center}
\includegraphics*[width=0.95\textwidth]{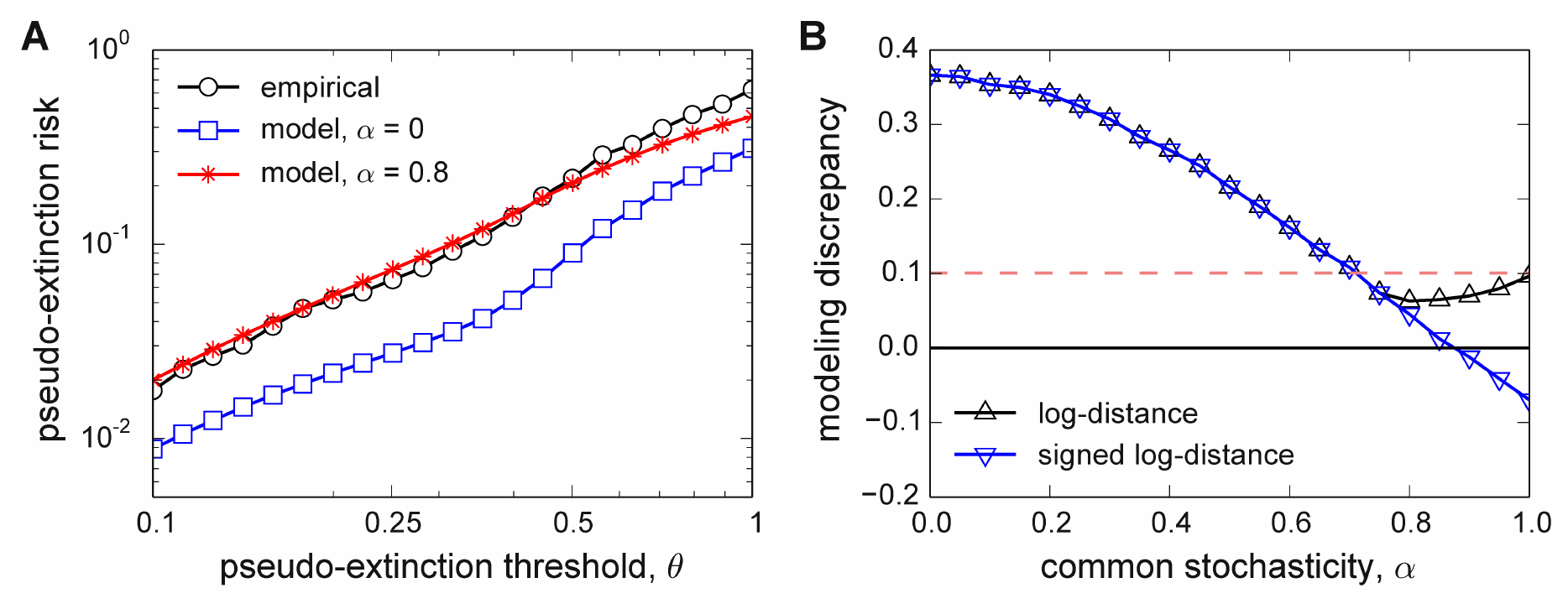} 
\end{center}
\caption{\label{common_stoch_fig}
\textbf{Model validation for ecosystem pseudo-extinction risk.}
\,
(\textbf{A})  Pseudo-extinction risk of the weighted average population of the ecosystem as a function of the threshold $\theta$: empirical data (circles),  model for $\alpha=0$ (squares), and model for $\alpha=0.8$ (stars), where $\alpha$ is the portion of the stochasticity common to all species in the corresponding ecosystem.  Each curve is an average over all ecosystems for the same 45-year parameter-determination period and 12-year prediction period used in figure~\ref{model_validation_fig}A. 
(\textbf{B}) Model discrepancy as a function of $\alpha$: average log-distance (upward triangles) and average signed log-distance (downward triangles) from the model curve to the empirical curve for the range of $\theta$ shown in panel~A. For comparison we highlight the limiting case of $\alpha = 1$ (dashed line), which is equivalent to applying our model directly to the weighted average population of each ecosystem.
}
\end{figure*}

\newpage
\clearpage
\renewcommand{\refname}{Supplementary References}

\renewcommand\figurename{{\small \bf Fig.}}
\renewcommand{\theequation}{S\arabic{equation}}

\newcommand{\sfigref}[1]{Fig.~\ref{#1}}
\renewcommand{\thefigure}{S\arabic{figure}}

\setcounter{section}{0}

\hfill {\large \em Regularity Underlies Erratic Population Abundances in Marine Ecosystems}

\hfill {\large J. Sun, S.P. Cornelius, J. Janssen, K.A. Gray \& A.E. Motter}
\bigskip
\bigskip

\noindent
\begin{center}
{\huge\bf Supplementary Material}
\end{center}
\doublespacing

\section{Effects of Non-Constant Catchability} \label{catchability_sec}
For a given species, the potential relationship between the reported catch (landing) in a given year, $x_t$, and the underlying population abundance, $n_t$, can be expressed in the general form
\begin{equation}
\frac{x_t}{n_t} = f(n_t),
\label{catch-eq}
\end{equation}
where $f$ is a function that then describes the \emph{catchability}, or catch per unit population. For example, in the classical Type I fishery model, $f(n_t) = q E$, where $q$ is catchability per unit effort and $E$ is total effort \cite{Ricker1940}. Here we focus on more general functions $f$ and study their effects on growth rates and population dynamics. We will make the assumption that catchability approaches a constant as population size grows large, i.e., $f(n_t) \rightarrow a$ as $n_t \rightarrow \infty$.  Under this assumption, we explore the following three mostly commonly-considered scenarios.\\

\noindent \textbf{Baseline.\ }  This is the simplest scenario, under which the amount caught is directly proportional to the actual population abundance (i.e., $f(n_t) = a$). Results reported in the main text were obtained under this assumption. In this case, the growth rates calculated from catch data are equal to the actual population growth rates since $x_{t+1}/x_t = n_{t+1}/n_t$, and hence the particular value of $a$ does not affect the growth rate distribution nor the predictions of pseudo-extinction risk made by our stochastic model. \\

\noindent \textbf{Hyperdepletion.\ } In this scenario, catchability \emph{decreases} at low population abundance, which is typically assumed to mean that $f(0) = 0$ and  $df/dn_t > 0$ for all $n_t \geq 0$\cite{Wilberg2009}. As a concrete example, we consider the following functional form:
\begin{equation}
f(n_t) = \frac{a n_t}{n_t + b},
\label{depletion}
\end{equation}
where $b > 0$ is a parameter. Note that we can use Eq.~\ref{catch-eq} to solve for population in terms of catch in this case as
\begin{equation}
n_t = \frac{x_t + \sqrt{x_t^2 + 4a b x_t}}{2 a}.
\end{equation} \\

\noindent \textbf{Hyperstability.\ } In this scenario, catchability \emph{increases} at low population abundance, which is typically taken to mean that $f(0) >0$ and $df/dn_t < 0$ for all $n_t  \geq 0$ \cite{Wilberg2009}. As a concrete example, we consider the following functional form:
\begin{equation}
f(n_t) = \frac{a +  \sqrt{a^2 + 4c /n_t}}{2},
\label{stability}
\end{equation}
where $c > 0$ is an additional parameter. From this we obtain that
\begin{equation}
n_t = \frac{x_t^2}{a x_t + c}. 
\end{equation} 

Figure \ref{catchability_schematic_fig} illustrates the relationship between catch and underlying population abundance for the three scenarios above. In our analysis, we use the parameter values $a = 0.1$, $b = 10^4$, and $c  = 10^2$. These values were chosen based on the typical scales  of the catch data in the LME dataset. Note that for the purposes of this section, we will calculate growth rates and train/validate our model based on the \emph{population abundances}, which are obtained by transforming the catch data in the LME dataset according to one of the three catchability relations defined above.

Figure \ref{catchability_hist_fig} shows the distribution of growth rate fluctuations under the hyperstability and hyperdepletion scenarios. In both scenarios, we see that the distribution is indistinguishable from the distribution of growth rate fluctuations in the baseline scenario, which we have shown to be Laplace (Fig.\ \ref{stats_fig} and Fig.\ \ref{gr_individual_fig}C-D). As such, Laplace statistics should still form the basis of our predictive stochastic model under other conceivable catchability scenarios. But how do the model's predictions of pseudo-extinction risk fare when based on population rather than reported catch?

Figure \ref{catchability_model_fig} shows the model-predicted pseudo-extinction risk as a function of pseudo-extinction threshold compared to the empirically-observed risk for the hyperstability and hyperdepletion scenarios. 
As for the baseline scenario of constant catchability (Fig.~\ref{model_validation_fig}A), the inclusion of the floor effect significantly improves our model's prediction of pseudo-extinction risk at low pseudo-extinction thresholds. However, in the case of hyperstability, both models (with and without floor effect) systematically underestimate pseudo-extinction risk. 

\section{Analysis of Stock Assessments} \label{ram_sec}

We use catch (landing) data for the analysis in the main text because it lends itself to high-quality statistics, being available for a large number of years and for a large number of species. Indeed, catch data is the {\it only} available indicator of the population abundance of most marine species. Nonetheless, one must be cautious in using reported landings as a direct proxy for population abundance, since there are factors that can affect catches other than changes in underlying population abundance, such as extreme weather events and changes in market demand or fishing effort. 

To address the possibility that these artefacts may have affected our results, we have repeated our statistical analysis on the RAM Legacy Stock Assessment Database \cite{Ricard2012}, 
which is the largest and most up-to-date catalogue of marine stock assessments available. For each assessed stock (comprising a specific species and geographical location), the relevant data consist of yearly estimates of the total biomass. These data integrate multiple independent sources of information beyond catch data, such as species-specific biological information and the results of research surveys, and are consequently regarded as more accurate estimates of population abundance. To obtain meaningful statistics and facilitate comparison to results in the main text, we focus on those assessments that have at least 30 consecutive years of data and that have total biomass estimates measured in units of tonnes. Out of the 331 assessments in the database, 199 satisfy these criteria.

Figure \ref{ram_data_fig} shows the statistics of the normalised growth rate fluctuations calculated for the populations in the RAM database. As shown, the central findings in the main text hold true. Namely, the normalised growth rate fluctuations follow a double-exponential (Laplace) distribution with exponent close to $\sqrt{2}$, both in the case when data from all assessments are pooled together (Fig.~\ref{ram_data_fig}A) and at the individual population level (Fig.~\ref{ram_data_fig}B-C). This analysis provides evidence that our results are not artefacts of biased sampling or observational error, but rather reflect the true statistical patterns of the underlying population dynamics.

\newpage
\clearpage
\renewcommand{\markboth}[2]{}

{

}

\newpage
\setcounter{figure}{0}

\begin{figure}[h]
\centering

\end{figure}

\begin{figure}[h!]
\centering
\includegraphics[scale=0.8]{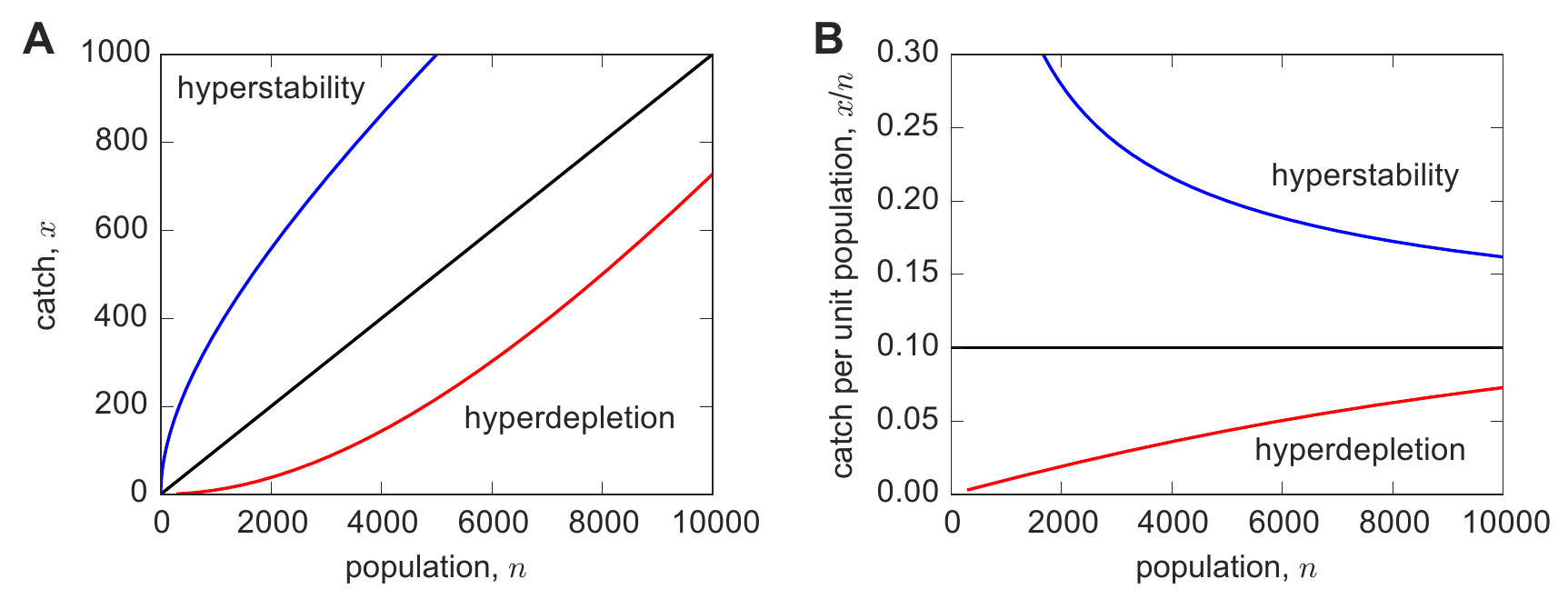}
\vspace{-0.2cm}
\caption{ \small \baselineskip14pt
{\bf Illustration of non-constant catchability.} 
(\textbf{A}) Relationship between underlying population abundance and catch under two different catchability scenarios: hyperstability (blue) and hyperdepletion (red),  in which catchability increases or decreases at low population abundances, respectively. These scenarios should be contrasted with the black curve, representing the assumption in the main text that catch is linearly related to abundance. (\textbf{B})
Corresponding catchability functions $f(n_t)$ described by Eq.~\ref{depletion} (red), Eq.~\ref{stability} (blue), and the constant catchability relation $f(n_t) = a$ (black). 
}\label{catchability_schematic_fig}
\end{figure}

\begin{figure}[h!]
\centering
\includegraphics[scale=0.8]{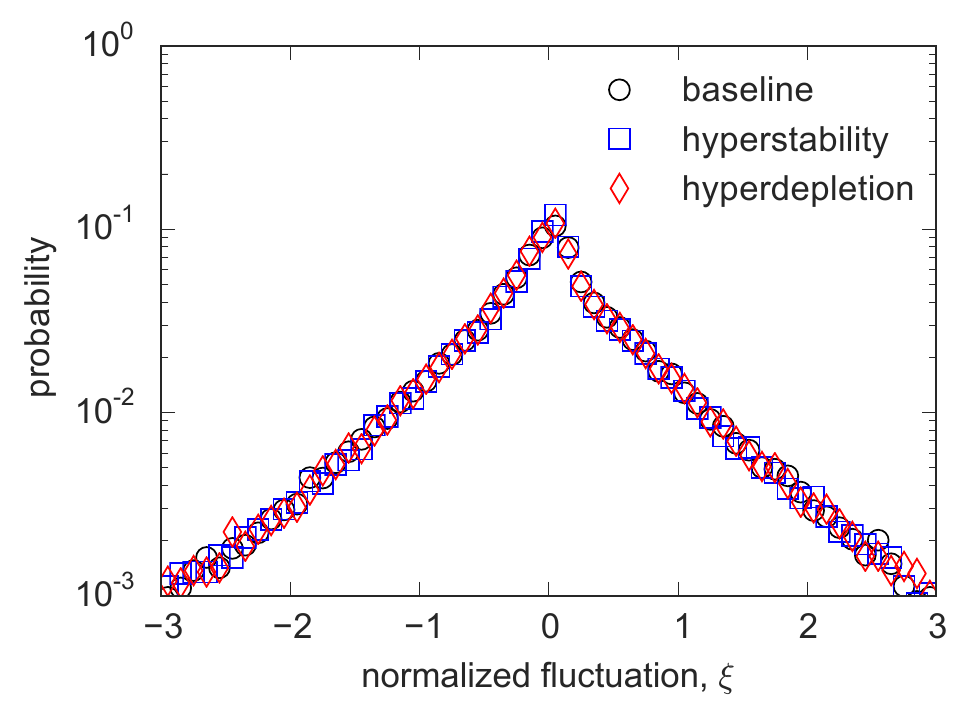}
\vspace{-0.2cm}
\caption{ \small \baselineskip14pt
{\bf Validation of the distribution of growth rate fluctuations under non-constant catchability.} Each curve represents the histogram of aggregated normalized growth fluctuations derived from the population abundances $\lbrace n^{(i)}_t \rbrace$, which are assumed to be related to the catch data $\lbrace x^{(i)}_t \rbrace$ in the LME dataset according to the given catchability relation: baseline, hyperstability, or hyperdepletion.
}\label{catchability_hist_fig}
\end{figure}

\begin{figure}[h!]
\centering
\includegraphics[scale=0.8]{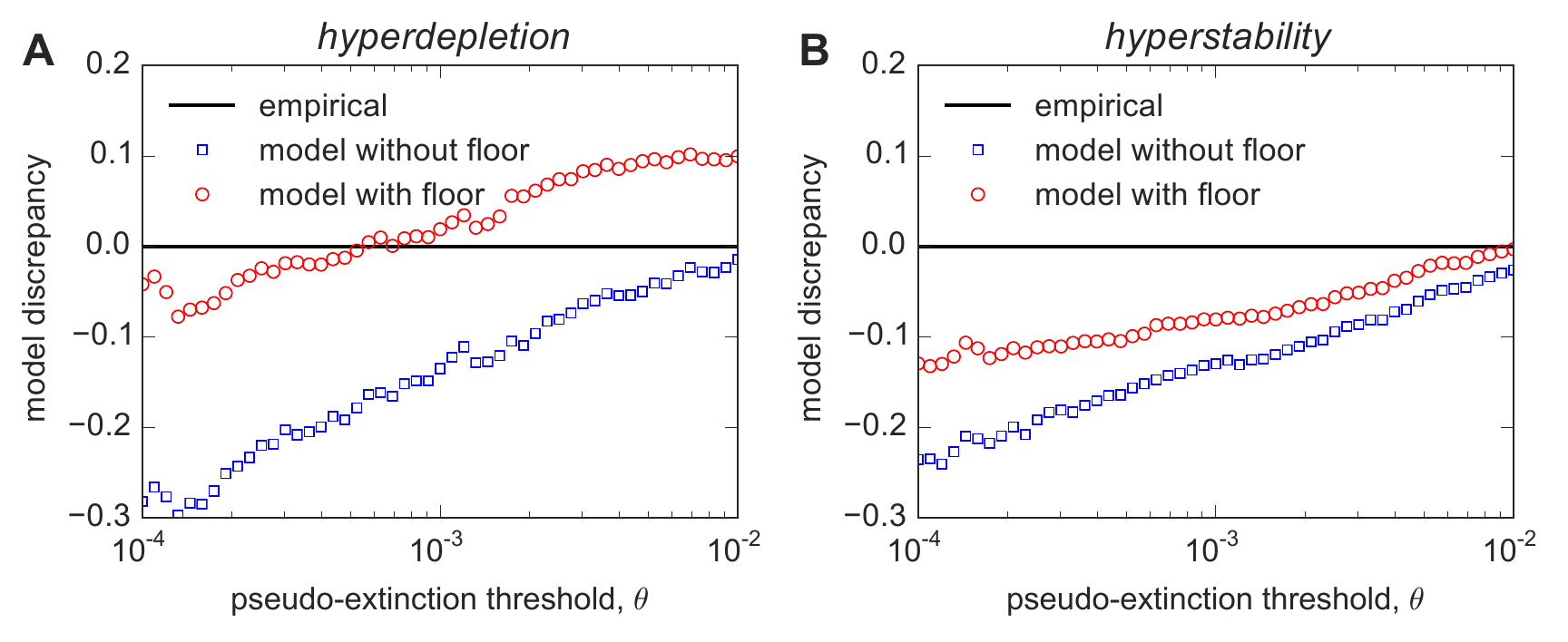}
\vspace{-0.2cm}
\caption{ \small \baselineskip14pt
{\bf Model validation for non-constant catchability.} Counterparts to Fig.~\ref{model_validation_fig}A for the hyperdepletion (left) and hyperstability (right) scenarios. 
}\label{catchability_model_fig}
\end{figure}

\begin{figure}[h!]
\centering
\includegraphics[scale = 0.55]{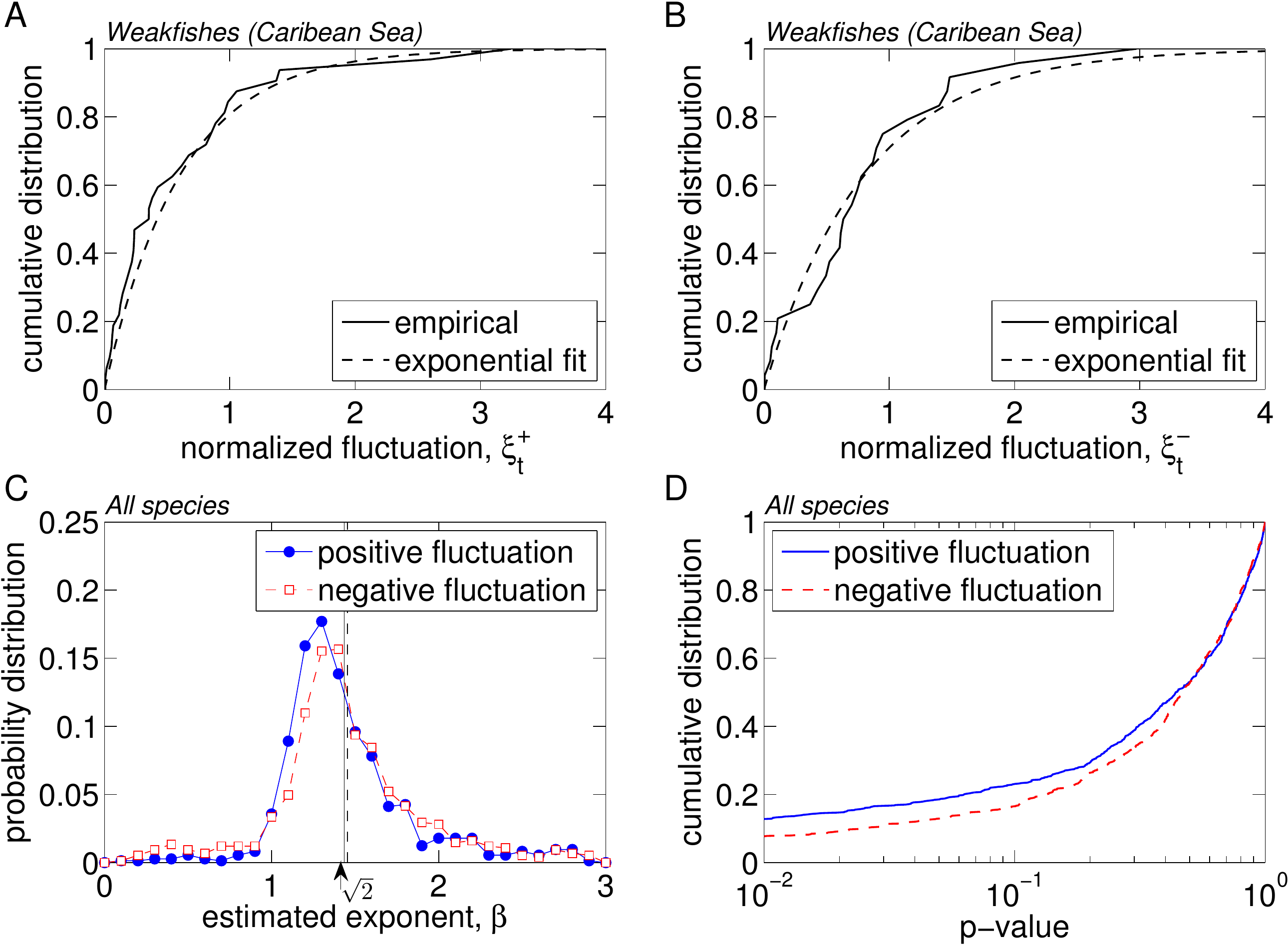} 
\vspace{-0.2cm}
\caption{ \small \baselineskip14pt
{\bf Statistics of the normalised growth rate fluctuations of {\em individual} species.} 
(\textbf{A} and \textbf{B}) Examples of the cumulative distributions for $\xi_t^+$ (positive fluctuations) and $\xi_t^-$ (negative fluctuations, in absolute value) along with the corresponding best exponential fits, where the exponent $\beta$ in $1- e^{-\beta \xi_t^{\pm}}$ is obtained by maximum likelihood estimation.  
(\textbf{C}) Distributions of the best fit exponents of the individual species, for all species in all 66 ecosystems under consideration.
The arrow indicates the exponent $\sqrt{2}$ obtained for the aggregate of all species (main text, Fig.~\ref{stats_fig}C), whereas the continuous and dashed lines indicate the mean exponent of individual species for positive and negative fluctuations, respectively. 
(\textbf{D}) Distributions of the {\it p}-values calculated using the Kolmogorov-Smirnov test between the empirical distributions of $\xi_t^\pm$ and their best exponential fits, represented in panel~\textbf{C}. Assuming, as usual, that the hypothesis is not rejected for {\it p}-values larger than $0.05$, it follows that the normalised growth rate fluctuations of most species are consistent with an exponential distribution with exponent close to $\sqrt{2}$. This confirms that the scaling for the aggregate data remains valid for individual species. We have verified that this conclusion also holds using the Akaike Information Criterion (AIC), which is another independent test of goodness of fit. According to the AIC, over $70\%$ of species' growth rate distributions more plausibly follow a Laplace distribution compared to the null hypothesis of a normal distribution.
}\label{gr_individual_fig}
\end{figure} 

\begin{figure}[h!]
\centering
\includegraphics[scale = 0.55]{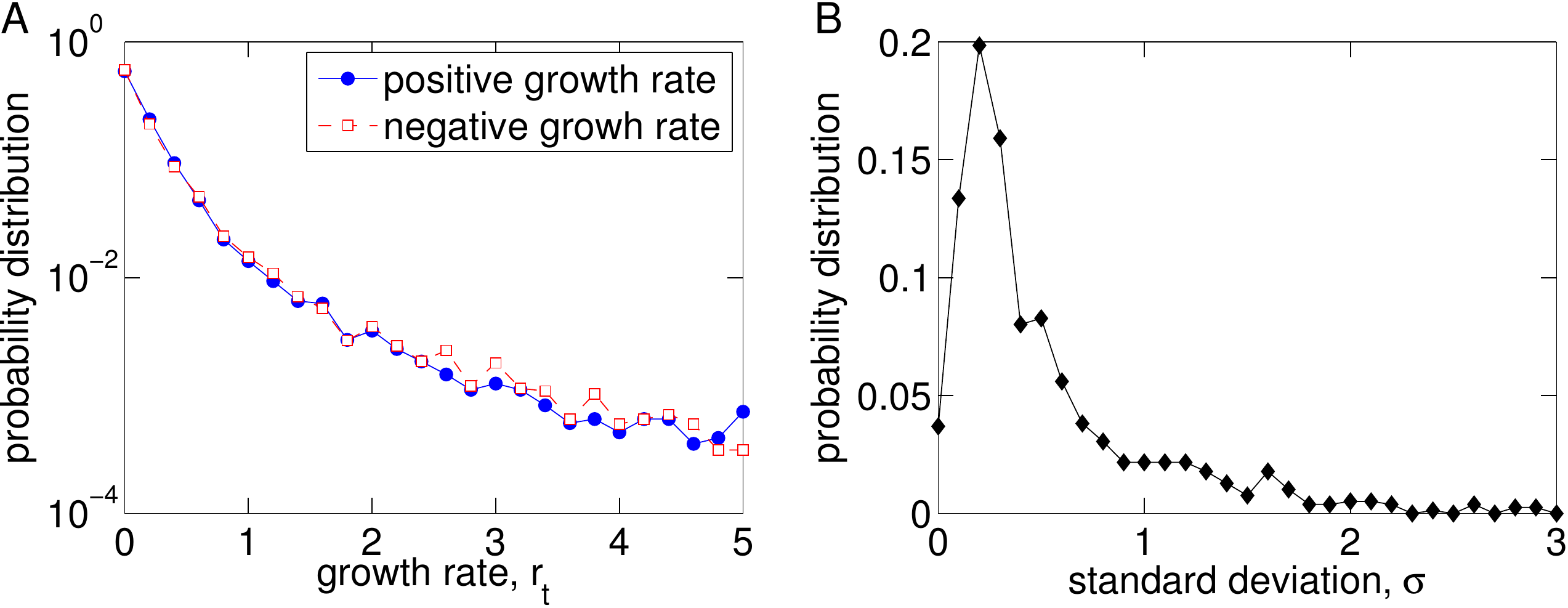}
\vspace{-0.2cm}
\caption{\small \baselineskip14pt
{\bf Statistics of {\em non-normalised} growth rates, $r_t$.}
(\textbf{A}) Non-normalised growth rates aggregated over all species, where the distributions of positive growth rates (solid symbols)
and negative growth rates (open symbols) are shown separately. 
This panel represents the non-normalised counterpart of the normalised growth rates considered in Fig.~\ref{stats_fig}C of the main text.
(\textbf{B}) Distribution across all species of each species' standard deviation of the growth rates,  $\sigma$. 
The distribution of non-normalised growth rates (panel A) has a heavier tail than the Laplace distribution representing
normalised growth rates; this difference is  due to heterogeneity across species, which is mainly reflected in the 
relatively broad distribution of the standard deviations (panel B). 
Thus, the normalisation introduced in our analysis 
is an important step  to reveal the scaling behaviour identified in this study.
}\label{gr_non_norm_fig}
\end{figure}

\begin{figure}[h!]
\centering
\includegraphics{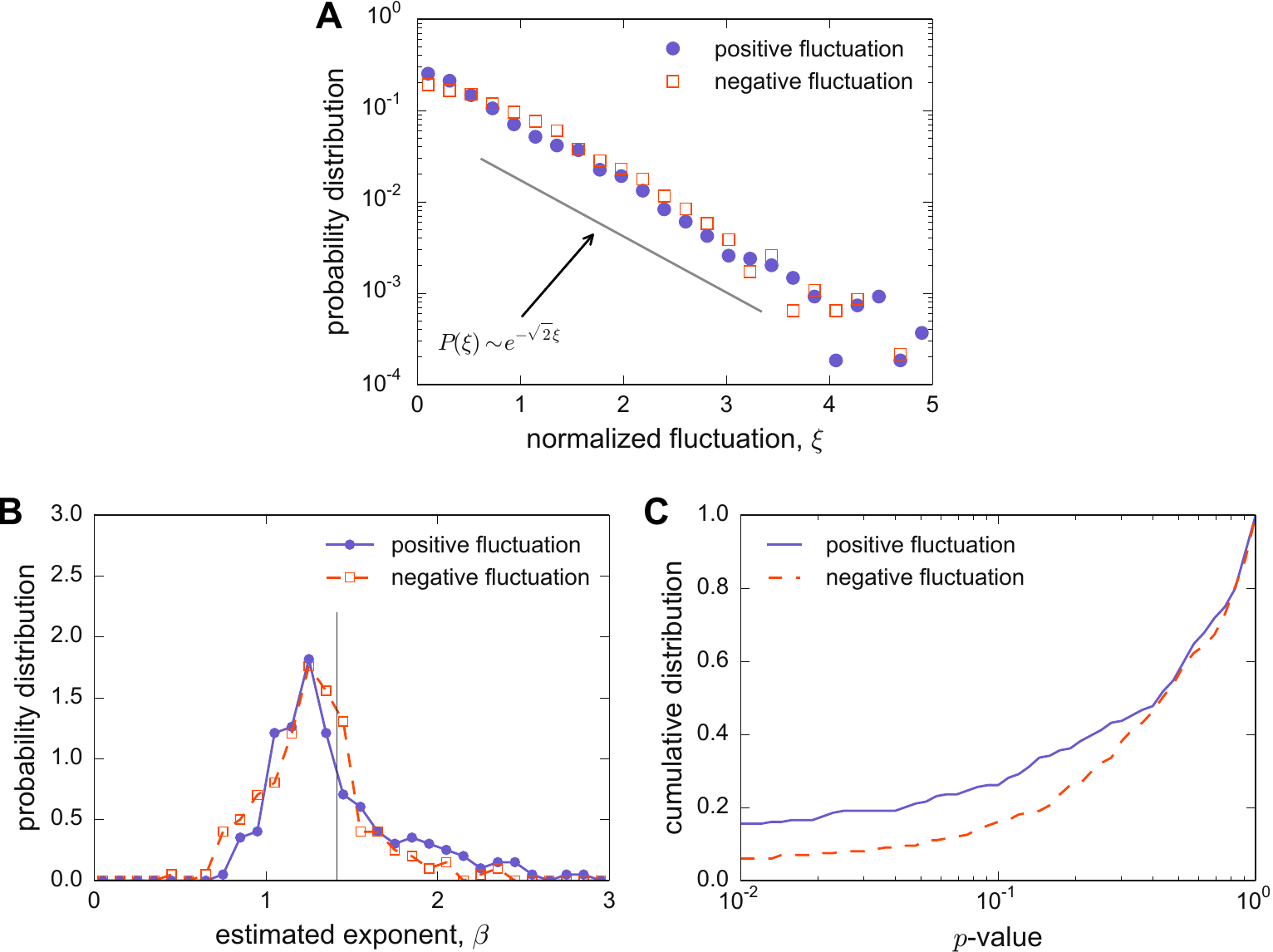}
\vspace{-0.2cm}
\caption{\small \baselineskip14pt
{\bf Statistics of normalised growth rate fluctuations in the RAM Legacy Stock Assessment Database.} (\textbf{A} to \textbf{C})
Counterparts to Fig.~\ref{stats_fig}C (main text), Fig.~\ref{gr_individual_fig}C, and Fig.~\ref{gr_individual_fig}D, respectively.
}\label{ram_data_fig}
\end{figure}

\clearpage
\newpage
\begin{figure}[h!]
\centering
\includegraphics[scale = 0.52]{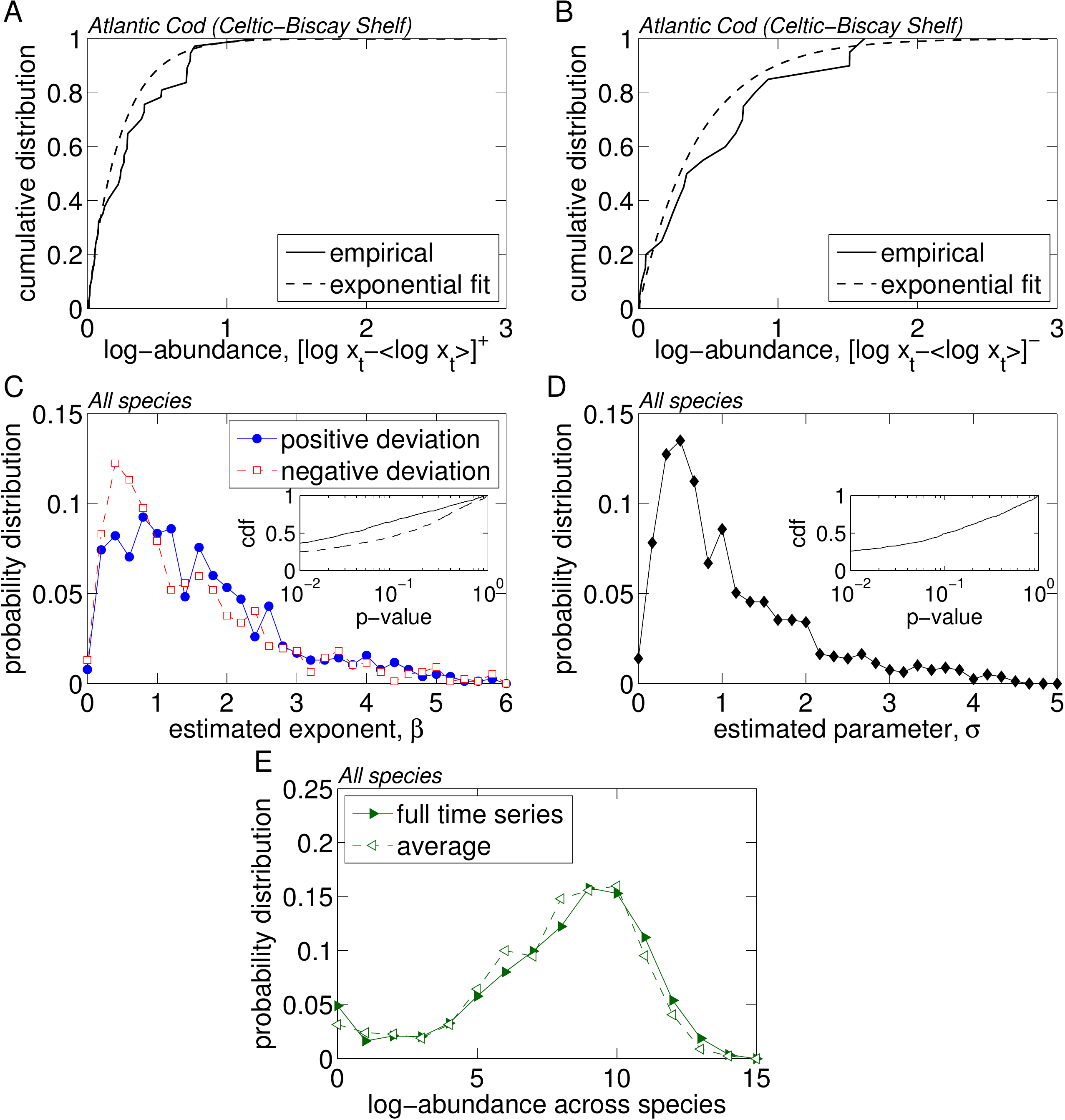} 
\vspace{-0.2cm}
\caption{\small \baselineskip14pt
{\bf Statistics of {\em population abundances}.}
(\textbf{A} and \textbf{B}) Examples of the cumulative distributions for $[\ln x_t -\langle\ln x_t \rangle]^+$ (positive deviation from the average) and $[\ln x_t -\langle\ln x_t \rangle]^-$ (negative deviation from the average, in absolute value) along with the best exponential fits for scaling exponents obtained by maximum likelihood estimation. These fits correspond to power-law probability distributions for the positive and negative deviations of the abundance itself, $P(x)=P_0x^{-1-\beta}$, where $P_0=\beta e^{\beta\langle \ln x\rangle}$.
(\textbf{C}) Distributions of the best fit exponents of the individual species, for all species 
under consideration. Inset: 
cumulative distributions of the corresponding {\it p}-values calculated using the Kolmogorov-Smirnov test, 
where the continuous (dashed) line corresponds to positive (negative) deviations. 
(\textbf{D}) Counterpart of panel C for best fits of  $\ln x_t $ by normal distributions, corresponding to lognormal distributions $P(x)=(x\sigma\sqrt{2\pi})^{-1}e^{\frac{-(\ln x -\langle\ln x \rangle )^2}{2\sigma^2}}$ for the abundances, where the standard deviations $\sigma =\sigma(\ln x_t)$ are determined by maximum likelihood estimation.
(\textbf{E}) Distributions of the average log-abundance per species,  $\langle\ln x_t \rangle$,  and 
of the full time-series of the log-abundances, $\ln x_t $, aggregated over all species. 
Panel E indicates that
 the {\it relative} species abundances are mainly determined by 
differences between the average populations of different species rather than by their time variations. On the other hand, the distributions of  the abundances of {\it individual species over time}
are generally no closer  to lognormal functions (inset of panel D) than they are to power-law functions (inset of panel C).
}\label{pop_statistics_fig}
\end{figure}

\end{document}